\documentclass[reprint,secnumarabic,amssymb, nobibnotes, aps, prd]{revtex4-2}
\usepackage{graphicx}
\usepackage{amsmath}
\usepackage{epstopdf}

\begin{document}

\title{Calculation of Kapitza resistance with kinetic equation}%

\author{A.\,P. Meilakhs} 
\email[A.\,P. Meilakhs: ]{mejlaxs@mail.ioffe.ru}
\author{ B.\,V. Semak} 
\affiliation{Ioffe Institute, 26 Politekhnicheskaya, St. Petersburg 194021, Russian Federation }
\date{\today}%

\begin{abstract}
A new method for calculation of interfacial thermal resistance in the case of heat transport through the interface by phonons is introduced in this research. The novelty of the suggested approach consists in the consideration of all the consequences of a non-equilibrium character of phonon distribution functions during the heat transfer. We use the well-described Diffuse Mismatch Model in order to introduce a model set of transmission and reflection amplitudes of phonons at the interface. An exact analytical solution for the proposed model is derived. Finally, the problem is solved for a set of transmission and reflection amplitudes characterized by a free parameter. 

\end{abstract}

\maketitle

\section{Introduction}
When heat flows through a boundary between two media a temperature jump occurs at the interface. A proportionality coefficient between the heat flux and the temperature jump is called  interfacial thermal resistance or Kapitza resistance \cite{Kap}. The inverse value called Kapitza conductance is also widely used. Investigation of the Kapitza resistance effect attracts a lot of attention in recent years for it is essential for the optimization of heat management of many applied systems including nanostructured materials for heat sink \cite{Appl1, Appl2}, nanofluids \cite{Appl3, Appl4}, thermoelectric generators \cite{Appl5, Appl6}, and nanoelectronics \cite{Appl7, Appl8}. 

Two most important models for the analytical description of heat transport through an interface are Acoustic Mismatch Model (AMM) and Diffuse Mismatch Model (DMM) \cite{Swar2}. In the first case, it is assumed that transmission and reflection amplitudes of phonons at the interface can be calculated with elasticity theory \cite{Kh, Can}. In the second case, it is assumed that interfacial scattering is very strong and the phonon incident on the interface "forgets"\ its initial direction and is scattered uniformly in all directions \cite{Swar}. In both models, the distribution function of the phonons incident on the boundary is assumed to be an equilibrium distribution function with the corresponding temperature. Meanwhile, the presence of constant heat flux across the interface means the presence of the same flux in the media, and, consequently, the phonon distribution functions differ from the equilibrium one.

Computer simulation of lattice dynamics at the interface is the most common approach to tackle the problems of Kapitza resistance theory described in the contemporary research manuscripts \cite{Din2}. Such questions as an effect of anharmonic scattering \cite{Din1} and lattice symmetry \cite{Din3} on phonon transmission through an interface, accuracy of DMM assumptions \cite{DinDMM}, a critical angle concept \cite{Din4}, and influence of chemical bonds on phonon transmission coefficient \cite{Din5} are investigated by computer modelling. These works significantly improve our understanding of the lattice dynamics at the interface; yet, questions about the form of the phonon distribution functions had not been previously considered. Papers on phonon kinetics at the interface \cite{PhonKin, PhonKin2} remain quite rare.

However, the questions of kinetics are fundamentally important for understanding the phonon heat transport through the interface. As it was first shown in Ref. \cite{Can},  formal application of the AMM formulae to calculation of the Kapitza resistance of some crystallographic plane in an ideal homogeneous crystal, produces a non-zero result. It is obvious that the correct calculation result should be exactly zero for such an imaginary boundary, on which the reflection of phonons does not occur. This paradox was resolved for the first time in Ref.\,\cite{Maas} where the one-dimensional model of two bound harmonic strings was used. In that research the authors applied the arguments quite similar to those used to describe electrical resistance of one-dimensional lattices \cite{Landauer}. It turned out that the source of the paradox is that AMM does not consider a non-equilibrium type of phonon distribution function. It is a clear manifestation that the kinetic assumption of the AMM (and also the DMM) is incorrect: the distribution function of phonons incident on the boundary is not an equilibrium distribution function with the corresponding temperature.

\begin{figure}[htbp]
\centerline{
\includegraphics[width=0.5\textwidth]{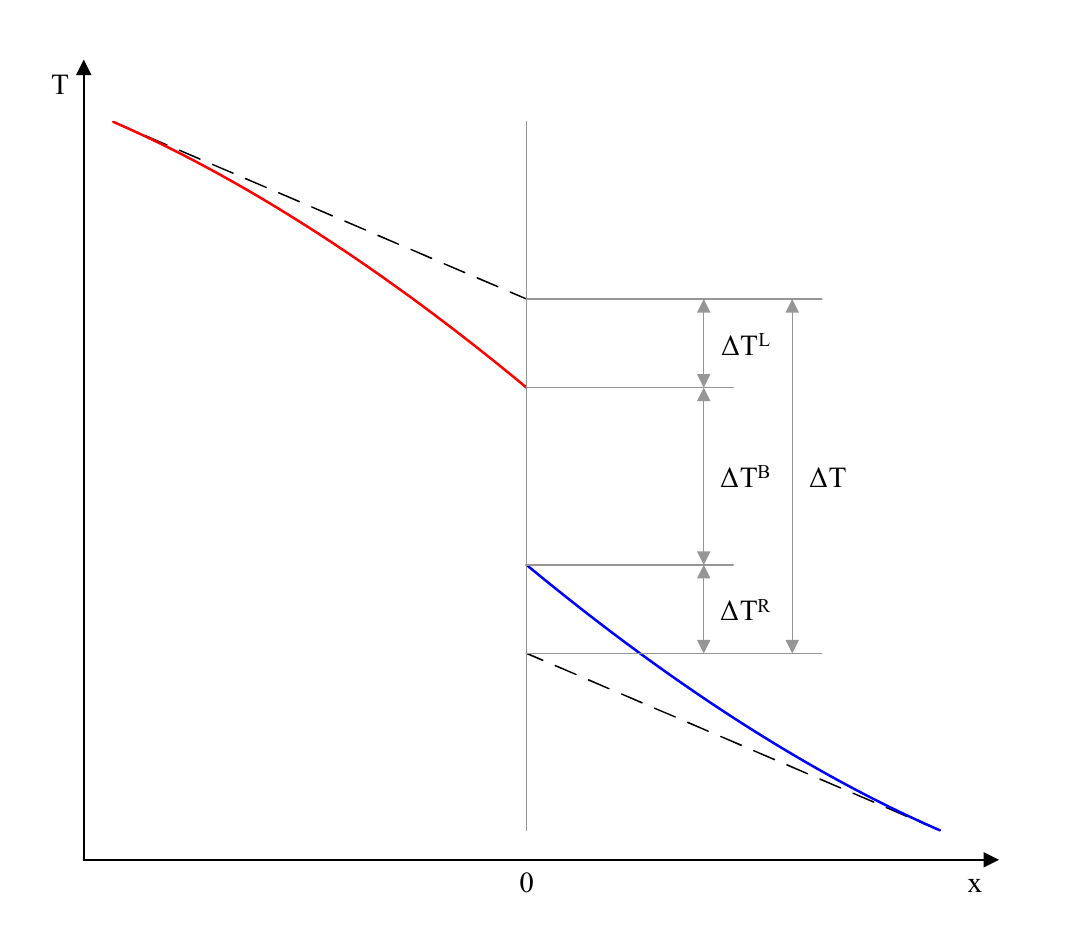}}
\caption{The temperature of the phonons near the interface with the heat flux from left to right. Shown are: the interface plane passing through zero along the coordinate $x$, solid curves show the course of the temperature of the phonons near the interface, the dashed straight lines show linear extrapolations of the course of the temperature to the interface. Denoted are: the temperature jump at the interface associated with the reflection of phonons at the interface $ \Delta T^B $;  the effective contributions to the temperature jump associated with smaller thermal conductivity near the interface, caused by interfacial scattering $\Delta T^{L, R}$;  the full temperature jump $ \Delta T $.} \label{fig1}
\end{figure}

On the other hand, in Ref. \cite{Mah} the question of the correct definition of the temperature of phonons on opposite sides of the interface is raised, given that the vibrational modes include the atoms of both crystals when the crystals are in contact: "Since the metal is bonded to the insulator, the phonon normal modes in equilibrium involve all of the atoms in both systems. However, during the heat flow, the phonons in the insulator are at a different temperature from those in the metal. This situation cannot be described by using the normal modes of the combined system."\, The definition of the temperature of phonons on opposite sides of the interface was proposed in \cite{Me}.

The defenition is based on the transition from the basis of combined modes to the basis of ordinary phonons, that is, those that correspond to an ordinary flat wave and only in one of the crystals. Wherein, the amplitudes of reflection and transmission of phonons at the interface are interpreted as the matching conditions for the distribution functions of phonons at the interface. With this approach, it is possible to create a formalism that naturally takes into account the nonequilibrium phonon distribution functions at the interface. For one-dimensional lattice this approach gives the same result as described previously \cite{Maas}.

Another manifestation of the non-equilibrium process of heat transport through the interface is an additional contribution to Kapitza resistance arising from the perturbation of the phonon distribution function by the interface \cite{Rel1}. Since the distribution function of phonons at the interface is perturbed by interfacial scattering, the distribution function of phonons near the interface is not equal to distribution function describing heat flow in a homogenous media. As it is known, the calculated thermal conductivity is maximal when it is calculated with the exact distribution function describing heat flow in the homogenous media \cite{LdKin}. Thus we can assume that thermal conductivity near the interface is smaller than in the homogenous media. And with the same heat flux, the temperature gradient at the interface is higher which leads to an additional effective contribution to the interfacial thermal resistance (Fig. 1). This contribution was investigated for various systems including the interface between normal and superconducting phases in type-II superconductors \cite{Rel2, Rel3}, interface between metal and insulator \cite{Rel4}, and dielectric nanolayers \cite{Rel5}. Since both contributions to Kapitza resistance, the one associated with the reflection of phonons at the interface and the one associated with smaller thermal conductivity near the interface, are crucially dependent on the form of the distribution function of phonons, they should be calculated simultaneously, within the framework of a unified system of equations.

In the present paper, the approach presented in Ref. \cite{Me} is generalized to the three-dimensional case. We introduce the full set of equations describing phonon kinetics at the interface. For the first time, we resolve the paradox mentioned in Ref. \cite{Can} for a three-dimensional system. We use the DMM-based set of transmission and reflection amplitudes of phonons at the interface, because of its popularity and simplicity. For this simple set of amplitudes, we find the exact solution. For the first time, we simultaneously calculate both contributions to Kapitza resistance, the one associated with the reflection of phonons at the interface and the one associated with smaller thermal conductivity near the interface. 

\section{System of equations}
In analytic consideration of the problem of the phonon transmission through the crystal boundary, both using the elasticity theory \cite{LdElast} and taking into account the atomic structure of the interface \cite {AnDin1, AnDin2, AnDin3}, the solution is sought, on the one side of the boundary, as a superposition of incident and reflected plane waves, and on the other side, in the form of a superposition of transmitted plane waves. The solution in this form, which includes oscillations of atoms on both sides of the interface, will be called the combined mode. Let us introduce the notation for the parameters of the combined modes.

Consider the interface of two semi-infinite crystals: the left one ($L$) and the right one ($R$). Let a unit amplitude wave, with a frequency $ \omega $, a component of the wave vector in the plane of the interface $ q_{||} $ and a polarization $j$, incident onto the interface from the left side. We denote as $\alpha$ the whole set of indexes characterising the wave: $\alpha = (q_{||},\omega, j)$.
The incident wave is partly reflected with an amplitudes $A_{\alpha \beta }^L$ and partly transmitted with an amplitudes $B_{\alpha \beta}^R$, where the first index stands for parameters of the incident wave and the second one for the parameters of the departing wave (Fig. 2a).

\begin{figure*}[htbp]
\centerline{
\includegraphics[width=0.9\textwidth]{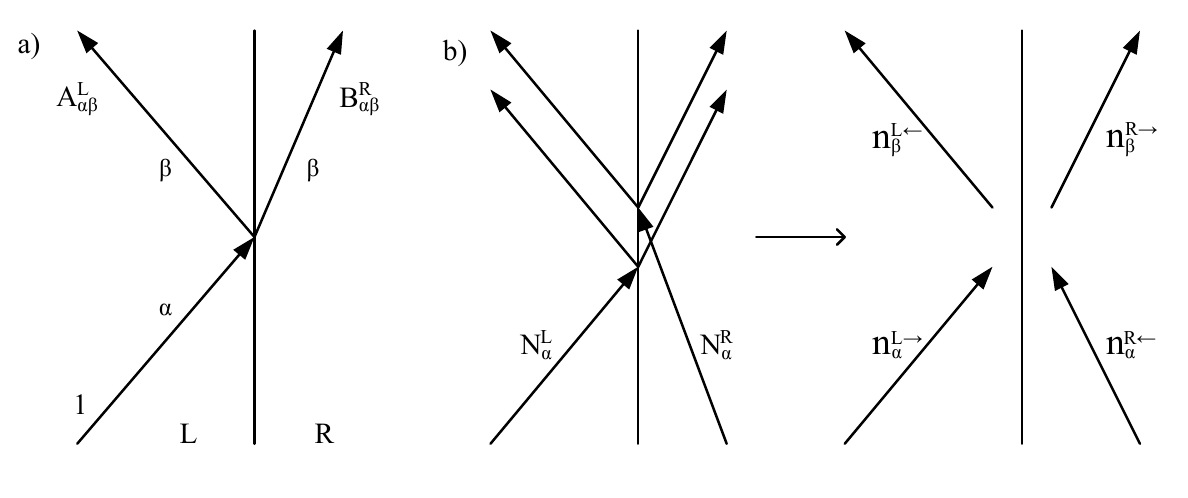}}
\caption{ a) Shown are: unit amplitude wave, incident onto the interface from the left side, with parameters $\alpha$, transmitted and reflected waves with parameters  $\beta$ and amplitudes $A_{\alpha \beta}$ and $B_{\alpha \beta}$ correspondingly. For simplicity the only one reflected and one transmitted wave are shown, while there can be infinitely many such waves because of the scattering. b) Shown is  the transition from the basis of combined modes with occupation numbers $N_{\alpha}^L, N_{\alpha}^R$ to the basis of ordinary flat waves with occupation numbers $n_{\alpha}^{L\rightarrow}, n_{\beta}^{L\leftarrow}, n_{\alpha}^{R\rightarrow}, n_{\beta}^{R\leftarrow}$. Indexes $\rightarrow, \leftarrow$ characterize the direction of the phonon movement, from left to right and right to left, respectively. Occupation numbers of departing phonons $n_{\beta}^{L\leftarrow}, n_{\alpha}^{R\rightarrow}$ are related to occupation numbers of incident phonons $n_{\alpha}^{L\rightarrow}, n_{\alpha}^{R\leftarrow}$ by distribution functions matching equations (2).} \label{fig2}
\end{figure*}

Whatever model of the phonon transmission across the interface we consider, the law of conservation of energy flow must be fulfilled for the combined mode. In other words, the energy flux to the interface and from the interface for each mode must be equal. Otherwise, the solution in the form of combined mode would not be stationary -- the interfacial atoms would have to increase or decrease the amplitude of oscillations over time. The energy flux in the direction perpendicular to the interface, $x$, is proportional to the wave velocity in the $x$ direction, the density of the substance, the square of the frequency, and the square of the wave amplitude. In this paper, we will not consider inelastic phonon scattering, so we can reduce the frequency. Thus, we have:
\begin{align} 
\rho^L v_{x,\alpha}^{L} = \rho^L \sum_{\beta } |A^L_{\alpha \beta }|^2 v_{x,\beta}^{L}  + \rho^R \sum_{\beta } |B^R_{\alpha \beta }|^2 v_{x,\beta}^{R}  \nonumber \\
\rho^R v_{x,\alpha}^{R} = \rho^R \sum_{\beta } |A^R_{\alpha \beta }|^2 v_{x,\beta}^{R} +  \rho^L \sum_{\beta } |B^L_{\alpha \beta }|^2 v_{x,\beta}^{L}. 
\end{align}
This property of the combined mode will be used later.

It is impossible to correctly determine the crystal temperatures using the basis of combined modes since the energy of each combined mode includes the vibrational energy of both crystals. Following the method proposed in \cite{Me}, we transit from the basis of the combined modes to the basis of ordinary plane waves (Fig. 2b). Herewith, the squares of the reflection and transmission amplitudes $A_{\alpha \beta }^{L,R}, B_{\alpha \beta}^{L,R}$ become the coefficients for the equation of matching of the distribution functions of phonons - plane waves. In the discrete case, these equations are written as:
\begin{align} 
n_{\alpha}^{L\leftarrow} =\sum_{\beta } |A^L_{\beta \alpha}|^2 n_{\beta}^{L\rightarrow} + \frac{\rho^L}{\rho^R}  \sum_{\beta } |B^L_{\beta \alpha}|^2 n_{\beta}^{R\leftarrow}  \nonumber \\
n_{\alpha}^{R\rightarrow} =\sum_{\beta } |A^R_{\beta \alpha}|^2 n_{\beta}^{R\leftarrow} + \frac{\rho^R}{\rho^L} \sum_{\beta } |B^R_{\beta \alpha}|^2 n_{\beta}^{L\rightarrow}. 
\end{align}

The physical sense of the matching equations (2) is very simple: phonons flying from the interface $n_{\alpha}^{L\leftarrow}, n_{\alpha}^{R\rightarrow}$ consist of those that fell on the interface from the same side and were reflected and of those fallen from the other side and transmitted $n_{\alpha}^{L\rightarrow}, n_{\alpha}^{R\leftarrow}$. Since the number of phonons in a given mode is proportional to the energy in a given mode, and the energy is proportional to the square of the amplitude, it is clear that it is the squares of the amplitudes that should be used as coefficients. The rigorous derivation of these equations is a straightforward generalization of the derivation of the matching equations in the one-dimensional case from Ref. \cite{Me}.

To describe the distribution function of phonons near the boundary, we use the conventional stationary Boltzmann equations in the relaxation time approximation:
\begin{equation}
v_x^{L,R}\frac{\partial n^{L,R}}{\partial x}=-\frac{\chi^{L,R}}{\tau^{L,R}}.
\end{equation}
Here $\chi^{L,R} = n^{L,R}-n_0^{L,R}$ is a nonequilibrium part of the distribution functions, $n_0$ -- is equilibrium part, that is Bose-Einstein distribution function.

Here we can introduce the definition of temperatures with phonon occupation numbers:
\begin{equation}
n_0^{L,R} = \frac{1}{\exp(\hbar\omega/k_B T^{L,R})+1}.
\end{equation}
In our method, we find $n_0^L, n_0^R$, and then $T^L, T^R$ is derived with equation (4). The naive method, that leads to the paradox described in the Introduction, defines $T^L, T^R$ by the same formula but with occupation numbers of combined modes  $N^L, N^R$. The elimination of the paradox with this changed definition of temperatures for the one-dimensional case is described in detail in Ref. \cite{Me}.

There is an ambiguity in the division of the distribution functions into the sum of the equilibrium and nonequilibrium parts. To eliminate this ambiguity, we introduce the Chapman-Enskog conditions \cite{LdKin}: the temperature of a nonequilibrium system is defined as the temperature of an equilibrium system with the same energy, which yields
\begin{equation}
  \sum_j \int \frac{d^3k}{(2\pi)^3}\chi^{L,R} \hbar\omega =0.
 \end{equation}

In order to connect the phonon distribution functions with the heat flux, we write
   \begin{equation}
  \sum_j \int \frac{d^3k}{(2\pi)^3}v_x\hbar\omega\chi^{L,R}=q.
   \end{equation}
Since the heat flux through the boundary is a conserved quantity, the last two equations are equivalent and can be restricted to an equation only for phonons in the left crystal. It is easy to verify that these equations are formally equivalent since the distribution functions obey the matching equations (2) where coefficients satisfy the flow conservation equation (1).

The matching equations for the phonon distribution functions (2) and the Boltzmann equations for phonons on both sides of the interface (3), together with the Enskog-Chapman conditions (5) and the expression for heat flux (6), make up the complete system of equations for  the phonon distribution functions on both sides of the interface.

It is easy to verify that the result of calculation of the Kapitza resistance for an imaginary boundary -- some plane in a homogeneous crystal, with the system of equations (2, 3, 5, 6) is exactly zero (unlike the AMM calculations). In this case
\begin{align} 
A_{\alpha \beta }^{L,R} &= 0  \nonumber \\
B_{\alpha \beta}^{L,R} &= \delta_{\alpha \beta}.
\end{align} 
For non-equilibrium distribution functions, we make a substitution that satisfies the Boltzmann equations (3):
\begin{equation}
\chi^{L, R} = -\tau^{L, R} v_x^{L, R} \frac{\partial n_0}{\partial T}  \frac{dT}{dx},\ \ \ \frac{dT}{dx} = q/\kappa.
   \end{equation}
And $\chi^L = \chi^R$ because in a homogeneous media $\tau^L = \tau^R = \tau$ and $v_x^L = v_x^R = v_x$.

Heat flux equation (6) is satisfied:
 \begin{equation}
    - \left( \sum_j \int \frac{d^3k}{(2\pi)^3} \hbar\omega \tau v^2_x \frac{\partial n_0}{\partial T} \right) \frac{dT}{dx} = q.
   \end{equation}
because the expression in parentheses is exactly the thermal conductivity in the relaxation time approximation $\kappa$. Chapman-Enskog conditions (5) are satisfied, since $v_x$ is an odd function of $k_x$ hence the integral on $k_x$ equals zero. Thus, we have found the non-equilibrium part of the distribution function.

 Substituting (7) into (2) we find that  $n_{\alpha}^{L\leftarrow} = n_{\beta}^{R\leftarrow}, n_{\alpha}^{R\rightarrow} = n_{\beta}^{L\rightarrow}$, and because  $\chi^L = \chi^R$ we find $n_0^L = n_0^R$. By definition (4) it meens $T^L = T^R = T$.

So distribution functions $n^L = n^R = n_0(T) -\tau v_x\frac{\partial n_0}{\partial T} \frac{q}{\kappa}$ are solutions of the system of equations (2, 3, 5, 6) with the transmission and reflection amplitudes defined by conditions (7).

Since $T^L = T^R$ there is no temperature jump associated with phonon reflection $\Delta T^B = 0$.  From expression (9) it can be seen that the temperature gradient at the interface is equal to the temperature gradient far from the interface $q/\kappa$, hence $\Delta T^L = \Delta T^R = 0$. The temperature jump on the imaginary boundary vanishes as it should be.

\section{Model}
Now that we derived a set of equations describing phonon kinetics at the interface in a general form, we want to formulate a specific model, i.e. a set of parameters which we want to substitue into equations (2, 3, 5, 6).

We use a Debye model of phonon spectrum. We also introduce the averaged values of the phonon velocities by the formula \cite{DebVel}:
   \begin{equation}
 \frac{3}{v^{L,R^2}}= \frac{2}{v^{L,R^2}_t}+ \frac{1}{v^{L,R^2}_l}.
   \end{equation}
Here $v^{L,R}_t$ is a transversal and $v^{L,R}_l$ is a longitudinal speed of sound in the corresponding crystal.

With this in mind, we rewrite the matching equations (2) in a more convenient continuous form and consider the reflection and transmission amplitudes as functions of angles, $\theta'$ -- angle of incidence, $\theta$ -- angle of reflection or transmission, angles are counted from the $x$ axis perpendicular to the interface:
   \begin{align}
    n^{L\leftarrow} = \int_0^1 dcos\theta' n^{L\rightarrow}(\theta^{'})  |A^L_{\theta\theta^{'}}|^2 +&  \nonumber \\ + \frac{\rho^Lv^{L^3}}{\rho^Rv^{R^3}} \int_0^1 dcos\theta' n^{R\leftarrow}(\theta^{'}) |B^L_{\theta\theta^{'}}|^2& \nonumber \\
    n^{R\rightarrow} = \int_0^1 dcos\theta' n^{R\leftarrow}(\theta^{'}) |A^R_{\theta\theta^{'}}|^2 +&  \nonumber \\ + \frac{\rho^Rv^{R^3}}{\rho^Lv^{L^3}}\int_0^1 dcos\theta' n^{L\rightarrow}(\theta^{'}) |B^R_{\theta\theta^{'}}|^2&.
   \end{align}
The ratio of velocity cubes before the second term in the right-hand sides arises as the ratio of the densities of states in the Debye model when transiting to the continuous limit.

We will now choose a set of phonon reflection and transmission amplitudes $A^{L,R}_{\theta\theta'}, B^{L,R}_{\theta\theta'}$. We take as a basis the simple for calculations and often used in the literature DMM \cite{Swar}. The choice of such a model also justifies the use of averaged velocities (10), since in this model the details of the phonon spectrum are not important, but only the phonon densities of states are important.

In DMM, it is assumed that the amplitudes of the reflected and transmitted waves do not depend on the angle of reflection or transmission, as well as on the angle of incidence of the wave on the boundary:
   \begin{align}
  |A^{L,R}_{\theta\theta'}|^2 =  |A^{L,R}|^2  \nonumber \\
|B^{L,R}_{\theta\theta'}|^2 =  |B^{L,R}|^2.
   \end{align}
However, we want to consider the coherent scattering of phonons at the interface (as, for example, in \cite{AnDin3}) and the description of the lattice dynamics with combined modes. It is easy to see that assumption (12) is not compatible with the condition of conservation of energy flow in each combined mode (1). Instead of the assumption of the classical DMM, we introduce the following assumption: the fraction of the energy flux dissipated in a certain direction does not depend on the angle of incidence of the wave on the interface. Energy flux of the incident wave in the direction perpendicular to the interface is proportional to  $v \cos \theta' $. So this assumption yields
   \begin{align}
|A^{L,R}_{\theta\theta'}|^2 = |A^{L,R}|^2 \cos \theta' \nonumber \\
|B^{L,R}_{\theta\theta'}|^2 = |B^{L,R}|^2 \cos \theta'.
   \end{align}
Let us find the set of amplitudes $A^{L,R}, B^{L,R}$.

When substituting the equilibrium distribution function into system (2), the system must become an identity for any set of reflection and transmission amplitudes, since at the same crystal temperatures there is no heat transfer between them, and the phonon distribution functions are equilibrium distribution functions.
   \begin{align}
    1 = \int_0^1 dcos\theta' |A^L_{\theta\theta'}|^2+\frac{\rho^Lv^{L^3}}{\rho^Rv^{R^3}} \int_0^1 dcos\theta' |B^L_{\theta\theta'}|^2 \nonumber \\
  1 = \int_0^1 dcos\theta' |A^R_{\theta\theta'}|^2+\frac{\rho^Rv^{R^3}}{\rho^Lv^{L^3}} \int_0^1 dcos\theta' |B^R_{\theta\theta'}|^2.
   \end{align}
This equation is analogous to the principle of the detailed equilibrium of the classical DMM \cite{Swar2}.

For the conservation of flow equation (1), we proceed to the continuous form in the same way as we did for the matching equations:
   \begin{align}   
    cos\theta^{'}=\int_0^1 dcos\theta cos\theta |A^L_{\theta\theta'}|^2 + \frac{\rho^Rv^R}{\rho^Lv^L} \int_0^1 dcos\theta cos\theta  |B^R_{\theta\theta'}|^2 \nonumber \\
    cos\theta^{'}=\int_0^1 dcos\theta cos\theta |A^R_{\theta\theta'}|^2+\frac{\rho^Lv^L}{\rho^Rv^R}\int_0^1 dcos\theta cos\theta |B^L_{\theta\theta'}|^2.
   \end{align}

We substitute the expression (13) into the equations (14, 15) and get
  \begin{align}   
   2=|A^L|^2 + \frac{\rho^L v^{L^3}}{\rho^R v^{R^3}} |B^L|^2 \nonumber \\
   2=|A^R|^2 + \frac{\rho^R v^{R^3}}{\rho^L v^{L^3}} |B^R|^2 \nonumber \\
   2=|A^L|^2 + \frac{\rho^R v^R}{\rho^L v^L} |B^R|^2 \nonumber \\
   2=|A^R|^2 + \frac{\rho^L v^L}{\rho^R v^R} |B^L|^2.
   \end{align}
The rank of this system of equations is 3. In order to obtain a complete system of equations for determining amplitudes $A^{L,R}, B^{L,R}$, we add the condition of the independence of the fraction of energy transmitted in a certain direction from the side of incidence $ (L, R) $, same as in DMM. Thus
  \begin{align}   
   |A^L|^2 = \frac{\rho^L v^L}{\rho^R v^R} |B^L|^2 \nonumber \\
   |A^R|^2 = \frac{\rho^R v^R}{\rho^L v^L} |B^R|^2.
   \end{align}

The rank of the system of equations (16, 17) is 4, so we can find
\begin{align}   
   |A^L|^2 &= \frac{2 v^{R^2}}{v^{R^2}+v^{L^2}} \nonumber \\
   |A^R|^2 &= \frac{2 v^{L^2}}{v^{R^2}+v^{L^2}}  \nonumber \\
   |B^L|^2 &=  \frac{2 v^{R^2}}{v^{R^2}+v^{L^2}}\frac{\rho^R v^R}{\rho^L v^L} \nonumber \\
   |B^R|^2 &=  \frac{2 v^{L^2}}{v^{R^2}+v^{L^2}} \frac{\rho^L v^L}{\rho^R v^R}.
   \end{align}
This is the desired model set of transmission and reflection amplitudes.

We will also assume that the phonon relaxation times are independent of the frequency $ \tau^{L, R} = const (\omega) $. This assumption is strictly fulfilled for strongly non-perfect \cite{Zaim} crystals, and is thus justified for areas near the boundary. In addition, as it will be seen later, the final answer will not depend on $ \tau^{L, R} $.

We will call the system of equations (2, 3, 5, 6) with amplitudes (18) and the Debye model of the phonon spectrum and the constant relaxation times "kinetic DMM".

\section{Solution}
Let us find the solution of kinetic DMM. We first find the distribution functions of the phonons incident on the interface by solving the stationary Boltzmann equation (3). Due to the perturbation of the distribution function of phonons by interfacial scattering, we must consider the non-equilibrium part of the distribution function as a function of the coordinate $\chi^{L,R}(x)$. For the left crystal
   \begin{equation}
   \frac{\partial \chi^L}{\partial x}+\frac{\chi^L}{v^L_x\tau^L}=-\frac{\partial n^L_0}{\partial x}.
   \end{equation}
This is an ordinary differential equation for  $x$, and general solution of the non-homogeneous equation is the sum of a particular solution and the complementary solution of the associated homogeneous equation, $\chi^L = \chi^L_p + \chi^L_c$. Complementary solution is:
   \begin{equation}
   \chi^L_c=\chi^L_0\exp{(-x/v^L_x\tau^L)},
   \end{equation}
where $\chi^L_0 = \chi^L(x=0)$. Similarly for the right crystal $\chi^R_c=\chi^R_0\exp{(x/v^R_x\tau^R)}$.

For incident phonons $v^L_x>0, v^R_x<0$, the solution of this type increases infinitely. Thus, it turns out that the complementary solution for the distribution function of incident phonons does not satisfy the boundedness condition, which means that the distribution function of incident phonons is determined only by a particular solution.

Particular solution is
\begin{equation}
   \chi^{L,R}_p =-\tau^{L,R} v^{L,R}_x\frac{\partial n^{L,R}_0}{\partial T} \left(\frac{dT}{dx} \right)^{L,R}.
   \end{equation}
Here  $\left(\frac{dT}{dx} \right)^{L,R}$ -- unknown functions of coordinate, since, due to the perturbation of the phonon distribution functions by the interface, the temperature gradients near the interface differ from the gradients in a homogeneous media for the same value of the heat flux.  So we have three unknown parameters, the temperature jump at the interface $\Delta T^B$, and temperature gradients on both sides of the interface. For these parameters to have the same dimension, instead of gradients we introduce:
\begin{equation}
  \tilde {\Delta  T^{L,R}} = \tau^{L,R} v^{L,R} \left(\frac{dT}{dx} \right) \Bigr|_{x=0}^{L,R}.
   \end{equation}
Because the considered problem is linear, we can assume the effective contributions to the temperature jump related to smaller thermal conductivity near the interface are proportional to temperature gradiens, so $\Delta T^{L,R} \sim  \tilde {\Delta  T^{L,R}}$.

 In the case of linear heat transport across the interface, temperature jump at the interface is much smaller than temperatures of media on both sides of the interface. We expand destribution function of phonons on one side of the interface  into  the Tailor series: 
   \begin{align}
\Delta T^B \ll  T =  T^R, \,  T^L = T + \Delta T^B \nonumber \\
 n_0(T^L) = n_0(T) +  \frac{\partial n_0}{\partial T}\Delta T^B.
   \end{align}
 We substitute the expression (22) into (21), and express the distribution functions of incedent phonons at the interface $ (x = 0) $  with parameters  $\tilde {\Delta  T^{L,R}}, \Delta T^B$:
   \begin{align}
n^{L\rightarrow} = n_0(T) + \frac{\partial n_0}{\partial T} \Delta T^B +  \frac{\partial n_0}{\partial T}\Delta \tilde T^L  \cos \theta' \nonumber \\
n^{R\leftarrow} = n_0(T) +  \frac{\partial n_0}{\partial T}\Delta \tilde T^R \cos \theta'.
   \end{align}
Here we also expanded $v_x = v \cos \theta'$.

Using the matching equations (2), we express the distribution functions of receding phonons with the distribution functions of incident phonons. We will use the matching equations in the form (11). Substituting (24) into (11) with amplitudes (18) we find
 \begin{align}
n^{L\leftarrow} =  n^{R\rightarrow} = n_0(T) + \frac{v^{R^2}}{v^{R^2}+v^{L^2}} \frac{\partial n_0}{\partial T} \Delta T^B +  \nonumber \\ + \frac{2}{3}\frac{v^{R^2}}{v^{R^2}+v^{L^2}} \frac{\partial n_0}{\partial T}\Delta \tilde T^L  + \frac{2}{3}\frac{v^{L^2}}{v^{R^2} + v^{L^2}} \frac{\partial n_0}{\partial T}\Delta \tilde T^R .
 \end{align}
The distribution functions of phonons near the boundary are thus expressed through three unknown parameters: $\Delta \tilde T^{L,R}, \Delta T^B$. 

In order to finally find the distribution functions of phonons near the boundary, we substitute (24, 25) into equations (5, 6) and obtain a system of three equations with three unknowns. We use $\chi^L = n^L - n_0(T^L)$. Here is the result of substitution into equation (6) for phonons in the left crystal:
   \begin{align}
& q= 3 \int \frac{d^2k_{||}}{(2\pi)^3}  \int \limits_0^{+\infty} dk_x v \cos \theta' \hbar\omega\frac{\partial n_0}{\partial T}\Delta \tilde T^L \cos \theta' - \nonumber \\
&- \int \frac{d^2k_{||}}{(2\pi)^3} \int \limits_{-\infty}^0 dk_x v \cos \theta \hbar\omega \Bigl[ \frac{v^{R^2}}{v^{R^2}+v^{L^2}} \frac{\partial n_0}{\partial T} \Delta T^B +  \nonumber \\ &+ \frac{2}{3}\frac{v^{R^2}}{v^{R^2}+v^{L^2}} \frac{\partial n_0}{\partial T}\Delta \tilde T^L  + \frac{2}{3}\frac{v^{L^2}}{v^{R^2}+v^{L^2}} \frac{\partial n_0}{\partial T}\Delta \tilde T^R -\frac{\partial n_0}{\partial T} \Delta T^B  \Bigr].
  \end{align}
Hear $k_{||}$ -- stands for the components of wave vector parallel to the interface.The multiplier 3 appears as a result of summation on three branches of the phonon spectrum.

We perform integration over angular components of integral, change the variable of integration from the wave vector to the frequency, and simplify the expression:
\begin{align}
q = \frac{3}{(2\pi)^2v^{L^2}}  \left( \int_0^{\omega_D}   \frac{\hbar^2}{k_B T^2} \frac{\exp(\hbar\omega/k_B T)}{(\exp(\hbar\omega/k_B T)+1)^2}  \omega^4 d\omega \right) * \nonumber \\ *
 \left[ \frac{1}{2} \frac{v^{L^2}}{v^{R^2}+v^{L^2}} \Delta T^B + \frac{1}{3}\frac{v^{L^2}}{v^{R^2}+v^{L^2}} \Delta \tilde T^L  - \frac{1}{3}\frac{v^{L^2}}{v^{R^2}+v^{L^2}}\Delta \tilde T^R \right] .
  \end{align}
Here $\omega_D$ is Debye frequency.

We make a substitution $z = \hbar \omega/k_B T$ for the integral. Now
\begin{align}
  \frac{\hbar^2}{k_B T^2}  \int_0^{\omega_D} \frac{\exp(\hbar\omega/k_B T)}{(\exp(\hbar\omega/k_B T)+1)^2}  \omega^4 d\omega =  \nonumber \\ = \frac{k_B^4 T^3}{\hbar^3} \int_0^{T_D/T} \frac{z^4 e^z}{(e^z-1)^2} dz.
   \end{align}
$T_D$ is Debye temperature in that of the crystals, in which the Debye temperature is lower. Let the Debye temperature be lower in the left crystal. 

One can see that the right part of the expression (28) is the Debye function. The Debuy temperature dependence naturally arises in the calculation of Kapitza conductance for the Debye model of phonon spectrum. However, later we will need this integral with a different lower limit. So we introduce the notation
\begin{equation}
 I_{T_1}^{T_2} = \int_{T_1/T}^{T_2/T} \frac{z^4 e^z}{(e^z-1)^2} dz.
   \end{equation}

We perform the same substitution for equations (5) and obtain
\begin{align}
- \frac{v^{L^2}}{v^{R^2}+v^{L^2}} \Delta T^B + \left(\frac{2}{3}\frac{v^{R^2}}{v^{R^2}+v^{L^2}}+\frac{1}{2}\right) \Delta \tilde T^L  +  \nonumber \\ + \frac{2}{3}\frac{v^{L^2}}{v^{R^2}+v^{L^2}}\Delta \tilde T^R = 0 \nonumber \\ 
 \frac{v^{R^2}}{v^{R^2}+v^{L^2}} \Delta T^B + \frac{2}{3} \frac{v^{R^2}}{v^{R^2}+v^{L^2}} \Delta \tilde T^L  +
  \nonumber \\ + \left(\frac{2}{3}\frac{v^{L^2}}{v^{R^2}+v^{L^2}}+\frac{1}{2}\right)\Delta \tilde T^R =0.
  \end{align}

Equations (27, 30) are system of three equations with three unknowns $\Delta T^B, \Delta \tilde T^{L,R}$. The solution is:
  \begin{equation}
   \Delta T^B =\frac{8 \pi^2}{7} \frac{\hbar^3 q}{k_B^4} \frac{v^{R^2}+v^{L^2}}{T^3 I_{0}^{T_D}},
   \end{equation}
and
  \begin{equation}
 \Delta \tilde T^{L,R} =\pm \frac{16 \pi^2}{7} \frac{\hbar^3 q}{k_B^4} \frac{v^{L,R^2}}{T^3 I_{0}^{T_D}}.
   \end{equation}
Note that $ \Delta \tilde T^R$ has a minus sign. 

We have found the phonon distribution function at the interface (24, 25) expressed with three parameters (31, 32). Now we want to investigate the relaxation of the distribution function of phonons disturbed by the interface to undisturbed function. That is a distribution function of phonons for constant heat flow in a homogenous media. So we will find the effective contributions to the temperature jump associated with smaller thermal conductivity near the interface $\Delta T^{L, R}$.

 To find the complementary part of the nonequilibrium distribution function at the interface, we use the expression
  \begin{equation}
\chi_c^L(0)=\chi_0^L = n^{L\leftarrow} -  n_0(T) - \frac{\partial n_0}{\partial T} \Delta T^B -  \frac{\partial n_0}{\partial T}\Delta \tilde T^L  \cos \theta'.
   \end{equation}
Substitute the expression for the complementary (20, 33) and particular (21) parts of the nonequilibrium distribution function in the heat flow expression (6) and find:
  \begin{equation}
 \frac{dT}{dx} =  - \frac{q}{\kappa} - \left( \frac{dT}{dx} \right)_{pert} ,
   \end{equation}
where $\left( \frac{dT}{dx} \right)_{pert}$ is a change in temperature gradient due to perturbation by the interface:
  \begin{equation}
 \left( \frac{dT}{dx} \right)_{pert} = \frac{3}{\kappa} \int \frac{d^3k}{(2\pi)^3} v^L_x \hbar\omega\chi_0^L \exp{(-x/v^L_x\tau^L)}.
   \end{equation}
Because of 
  \begin{equation}
\Delta T^{L} = -\int_{-\infty}^0 dx \left( \frac{dT}{dx} \right)_{pert},
   \end{equation}
after substituting relation (35) into (36) and integrating, we find 
 \begin{equation}
\Delta T^L = \frac{1}{8} \Delta \tilde T^L.
   \end{equation}
The calculation of $ \Delta T^R $ is performed similarly, with the difference that for phonons with frequencies greater than the Debye frequency in the left crystal, coefficients in the matching equation (18) should be changed so that $ B^L = 0 $, since such phonons do not pass through the interface. We obtain
  \begin{equation}
\Delta T^R = -\frac{1}{8} \Delta \tilde  T^R - \frac{1}{4} \left( I_{T^L_D}^{T^R_D}/I_{0}^{T^R_D} \right) \Delta \tilde T^R,
   \end{equation}
where the second term is the relaxation of the distribution function of those phonons that are not included in the heat transport through the interface. Since $\Delta \tilde T^R$ is a negative quantity, $\Delta T^R$ is positive.

\section{Results and discussion}

\begin{figure}[htbp]
\centerline{
\includegraphics[width=0.65\textwidth]{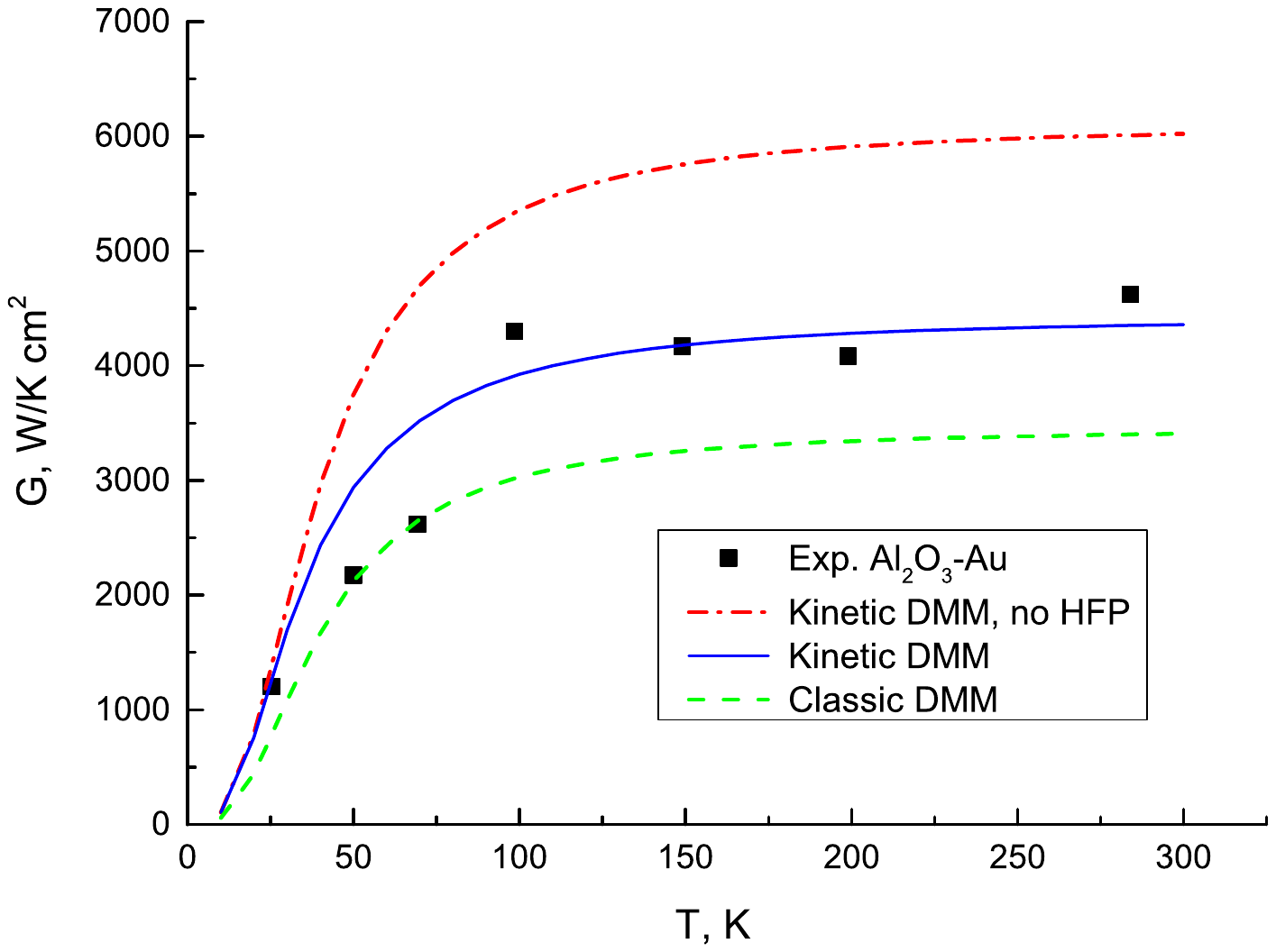}}
\caption{The dependence of the Kapitza conductance at the boundary of sapphire and gold on temperature. The points are experimental data from Ref. \cite{exp1}. The solid line is the result of calculations with kinetic DMM, formula (32). The dashed line is  the result of calculations with classic DMM. The dash-dotted line is kinetic DMM without taking into account the contribution of the relaxation of high-frequency phonons (HFP). Data is presented on a normal scale.} \label{fig3}
\end{figure}

 We have found all the contributions to temperature jump at the interface (31, 32, 37, 38). Now, the Kapitza conductance is expressed as
  \begin{align}
    G = \frac{q}{\Delta T^B+\Delta T^L+\Delta T^R} =  \nonumber \\ = \frac{7}{2 \pi^2} \frac{k_B^4} {\hbar^3}  \frac{T^3 I_{0}^{T_D}}{5 (v^{R^2}+v^{L^2}) + 2 v^{R^2} \left(I_{T^L_D}^{T^R_D}/I_{0}^{T^R_D}\right)}.
   \end{align}
It is interesting that, just like in classic DMM, in kinetic DMM, the Kapitza conductance turns out to be independent of the density of crystals. Also, from the expression (39) it is clear why the averaging of the velocities should be carried out according to the formula (10).

\begin{figure}[htbp]
\centerline{
\includegraphics[width=0.65\textwidth]{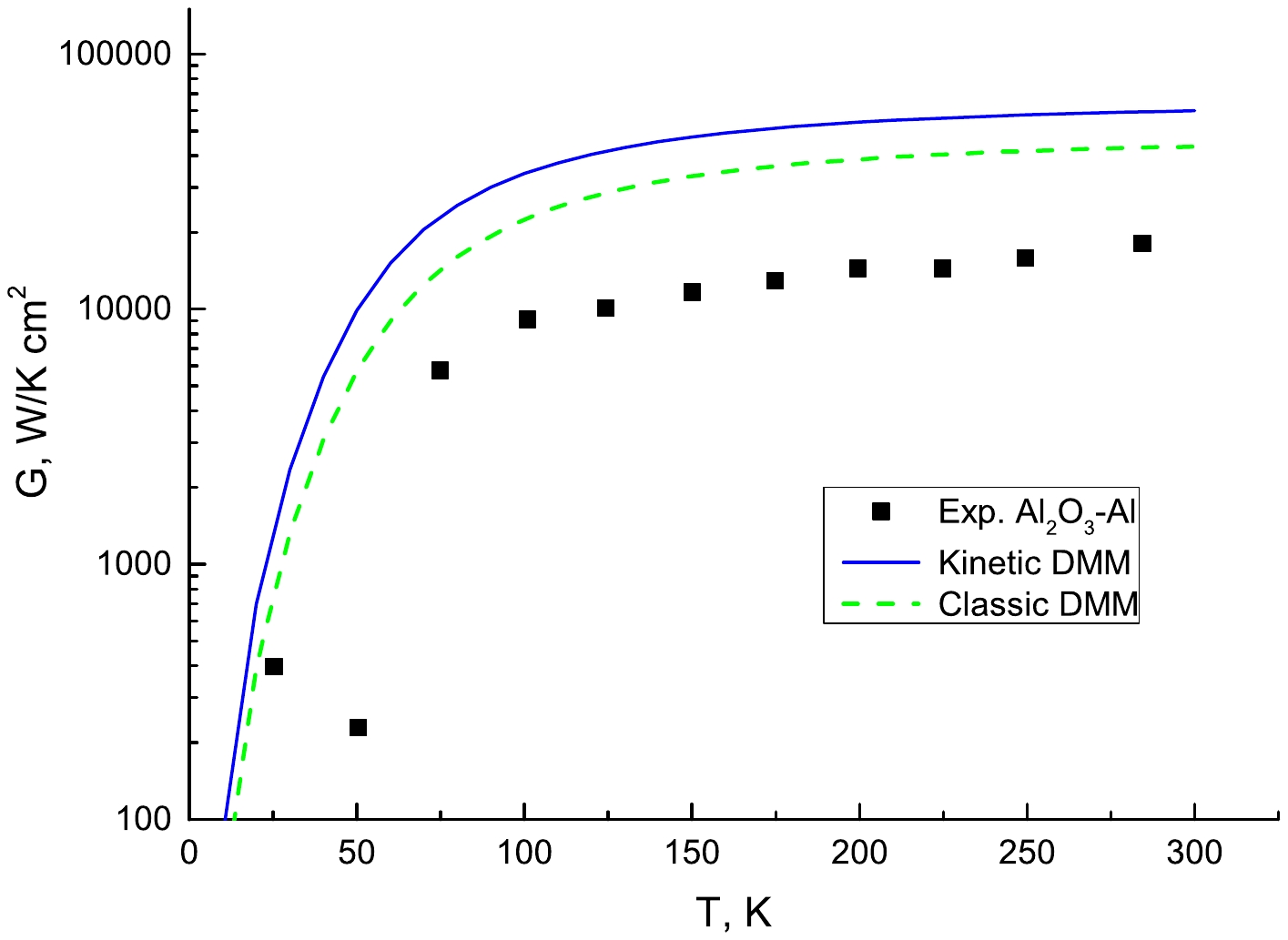}}
\caption{The dependence of the Kapitza conductance at the boundary of sapphire and aluminum on temperature. The solid line is the result of calculations with kinetic DMM, formula (39). The dashed line is  the result of calculations with classic DMM. The points are experimental data from Ref. \cite{exp1}. Data is presented on a logarithmic scale.} \label{fig4}
\end{figure}

The contribution of the relaxation of high-frequency phonons $2v^{R^2}\left(I_{T^L_D}^{T^R_D}/I_{0}^{T^R_D}\right)$ is small for a pair of substances with approximately equal values of the Debye temperature. However, if the Debye temperatures of two substances are significantly different, this contribution can be quite significant, given that the speed of sound is usually higher in the substance where the Debye temperature is higher. Considering the contribution from the relaxation of high-frequency phonons significantly improves the coincidence in the case when the results of kinetic DMM coincide well with the experimental results (Fig. 3).
 
 By removal the relaxation of high-frequency phonons factor (HFP) from the expression (39), we obtain the formula, which differs from the classical DMM formula, when averaging (10) is applied to, only by a numerical factor. The result of calculations using kinetic DMM without HFP is approximately $ 1.8 $ times greater. In many cases, classical DMM tends to overestimate the value of the Kapitza conductance. Kinetic DMM has an even worse agreement with experiment in such cases (Fig. 4).

\begin{figure}[htbp]
\centerline{
\includegraphics[width=0.48\textwidth]{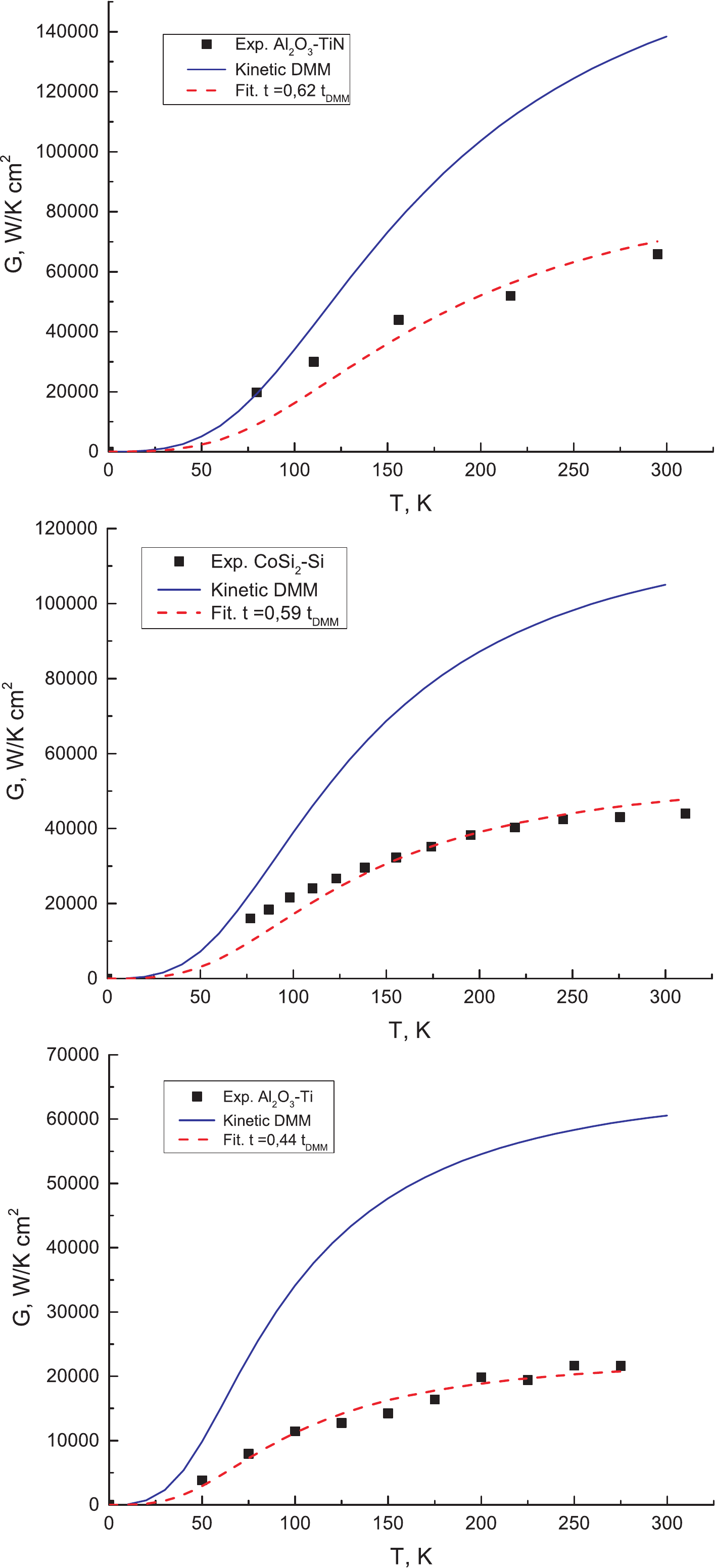}}
\caption{The dependence of the Kapitza conductance on temperature. The solid line is the result of calculations with kinetic DMM, formula (39). The dashed line is the result of fitting with the free parameter $t$ -- the transmission coefficient, formula (41). Data is presented on a normal scale. a)  The boundary of sapphire and titanium nitride. The points are experimental data from Ref.  \cite{exp2}. The optimal value of the free parameter is $t=0,62 t_{DMM}$. b)  The boundary of cobalt disilicide and silicon. The points are experimental data from Ref. \cite{exp3}. The optimal value of the free parameter is $t=0,59 t_{DMM}$. c) The boundary of sapphire and titan. The points are experimental data from \cite{exp1}. The optimal value of the free parameter is $t=0,44 t_{DMM}$.} \label{fig5}
\end{figure}

For a better understanding of the proposed method and model, we also consider a model with a free parameter. In order to introduce a free parameter, we discard a rather arbitrary condition of independence of the fraction of the energy transmitted in a certain direction from the side of phonon incidence (17) and introduce the transmission coefficient $t=|B^R|^2$. Having performed the calculations for this model, similar to those above, we find
 \begin{equation}
    G^B = \frac{q}{\Delta T^B} = \frac{3}{8 \pi^2} \frac{k_B^4} {\hbar^3}  \frac{T^3 I_{0}^{T_D} t}{\frac{\rho^L v^L}{\rho^R v^R}v^{L^2} - \frac{4}{7} (v^{R^2}+v^{L^2})t }.
   \end{equation}
We can see that the dependence on the density of crystals appeared in the new expression. Thus, it turns out that the independence of the Kapitza conductance (39) on densities was obtained due to the acceptance of assumption (17). Unexpectedly, the magnitudes of temperature jumps associated with smaller thermal conductivity near the interface $\Delta T^{L, R}$ are independent of $t$ and are determined by expressions (32, 37, 38).

Expression (40) equals to (39) at $t = \frac{\rho^L v^L}{\rho^R v^R} \frac{v^{L^2}}{v^{R^2}+v^{L^2}}$. We will call this value $t_{DMM}$. Let us express $G^B$ in terms of $t_{DMM}$
 \begin{equation}
    G^B =  \frac{3}{8 \pi^2} \frac{k_B^4} {\hbar^3}  \frac{T^3 I_{0}^{T_D} t}{ (v^{R^2}+v^{L^2})( t_{DMM} - \frac{4}{7}t) }.
   \end{equation}
It can be seen that the resistance associated with the reflection of phonons from the boundary goes to infinity, or, equivalently, the temperature jump at the boundary vanishes at $ t = 7/4 t_{DMM} $. The calculated dependency on the parameter $t$ is quite similar to the one calculated in Ref. \cite{Me}, which is $t/(1-t)$. Because in this paper we consider the more complex problem, the value 1 is  replaced with a quantity  $ 7/4 t_{DMM} $ that characterizes the chosen pair of materials. 

One can easily see that introducing the parameter $t$ to classic DMM in the same manner leads to linear dependence of Kapitza conductance on $t$. The non-linear dependence is a feature of the presented method that yields a zero value of Kapitza resistance at the imaginary boundary.  This feature is shown in the singular nature of the formulae (40, 41).

\begin{figure}[htbp]
\centerline{
\includegraphics[width=0.65\textwidth]{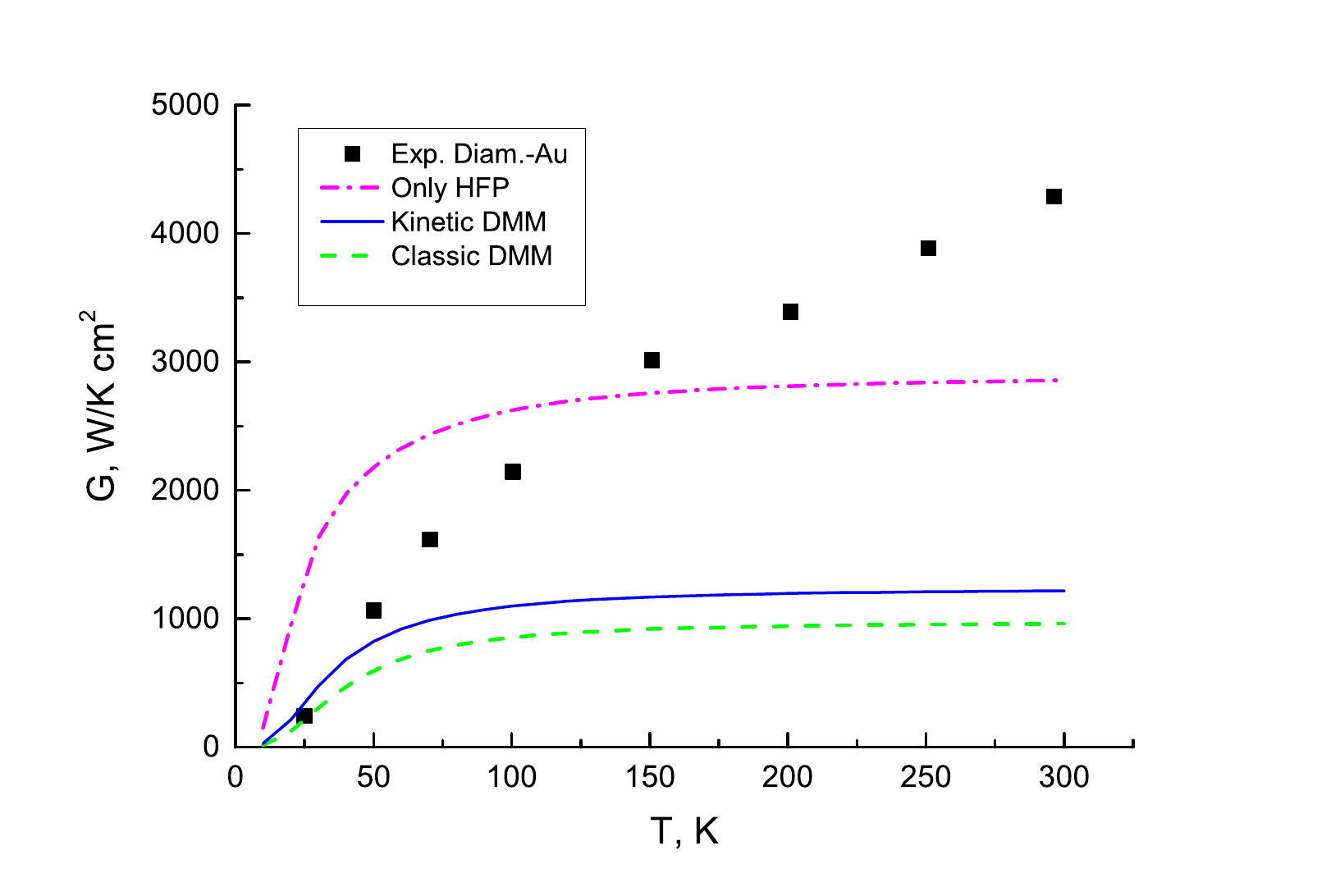}}
\caption{The dependence of the Kapitza conductance at the boundary of diamond and gold on temperature. The solid line is the result of calculations with kinetic DMM, formula (39). The dashed line is  the result of calculations with classic DMM. The dash-dotted line is kinetic DMM at $ t = 7/4 t_{DMM} $, thus taking into account only the relaxation contribution. The points are experimental data from Ref. \cite{exp1}. Data is presented on a normal scale.} \label{fig6}
\end{figure}

The nonlinear dependence of Kapitza conductance $G^B $ on the transmission coefficient $t$ (41), is a possible cause of the substantial discrepancy between the theory and an experiment in some cases.  For example, at $t = 3/4 t_{DMM}$, the conductivity decreases by almost two times $G^B (3/4 t_{DMM}) = 9/16 G^B (t_{DMM})$. Figure (5) shows the dependencies of $G(T)$ on the temperature at $t$ which gives the best approximation of the experimental data. It turns out that for most pairs of boundaries a good approximation is achieved by varying $t$ within $60 \% \, t_{DMM}$.

 The contribution of the relaxation of high-frequency phonons to the calculated value of Kapitza resistance is small at the low temperatures and is growing when the temperature grows. That leads to a slower growth rate of the overall Kapitza conductance with temperature. In some cases, it leads to a very good coincidence with the experimentally measured temperature dependence of Kapitza conductance (Fig. 3, Fig. 5c). However, in some cases it can be seen (Fig. 5a, Fig. 5b) that in the low-temperature region the theoretical curves pass below the experimental points, but at the temperatures close to room temperature the theoretical curves turn out to be, on the contrary, higher. We can offer two possible explanations of this deviation of experimental data from the temperature dependence predicted by the theory. It can be assumed that the DMM approach is not correct, the scattering at the boundary is rather weak, and in this case, as shown by theoretical models \cite{AnDin1, AnDin3}, the phonon transmission coefficient should decrease with increasing phonon frequency. This would explain the slower increase in the Kapitza conductance with increasing temperature, at which high-frequency oscillations are excited. It is possible, on the contrary, to assume that the DMM assumption about the diffuse nature of scattering is correct, and the discrepancy between theory and experiment is associated with the use of the Debye approximation. In Ref. \cite{DMMdisp} the Kapitza conductance was studied in the DMM model, but with a more realistic dependence of the density of phonon states on the frequency than that in the Debye model. As a result, it turns out that at low frequencies, Kapitza conductance grows faster with increasing temperature than classical DMM predicts, and at high temperatures, on the contrary,  it grows more slowly, which explains the observed discrepancy.

 In the framework of kinetic DMM we can not introduce such a concept as "Radiation limit"\ \cite{RadLim} since, as it was shown in a previous section, the temperature jump on the imaginary boundary vanishes. However the magnitudes of temperature jumps associated with smaller thermal conductivity near the interface $\Delta T^{L, R}$ are independant of $t$ so we can introduce the "Relaxation limit"\ which is kinetic DMM at $ t = 7/4 t_{DMM} $ so the temperature jump associated with the reflection of phonons at the interface vanishes, thus taking into account the only the relaxation contribution. The kinetic DMM, as well as classic DMM, predicts a significantly lower Kapitza conductance value than that measured in the experiment for the pairs of substances with very different sound velocities, such as, for example, diamond - lead, and diamond - gold (Fig. 6). In such cases, experimental data cannot be fitted by variation of the free parameter $t$, because even at "Relaxation limit"\ the interfacial thermal resistance  associated with smaller thermal conductivity near the interface is still larger than the experimentally measured. We share the common point of view \cite {Mah, ElPhon1} that this discrepancy can be explained only with the help of a model that considers the direct transfer of heat from the metal electrons to the dielectric phonons \cite{ElPhon2, ElPhon3, ElPhon4}, or a model that takes into account both phonon heat transport and electron-phonon coupling \cite{Both1, Both2}.

\section{Conclusion}
A new method for calculating the Kapitza conductance was proposed. Matching equations for the phonon distribution functions are introduced, which, together with the Boltzmann equations for phonons and some additional equations, form a complete system of equations (2, 3, 5, 6) describing the kinetics of heat transfer across the boundary by phonons. A method of solution to such a system was introduced. It is shown that the use of this method does not lead to a paradox: the solution of the corresponding system for an imaginary boundary, that is, for a plane in a homogeneous crystal, gives zero Kapitza resistance.  For a model set of amplitudes based on the classical DMM, an exact analytical solution is obtained. Moreover, the algorithm for solving the equations (2, 3, 5, 6) which can be applied to any set of transmission and reflection amplitudes is found. The dependence of Kapitza conductance on temperature is found.  The problem was also solved for a set of amplitudes with a free parameter characterizing the fraction of energy transmitted by a phonon across the boundary. A feature of the method that yields a zero value of Kapitza resistance at the imaginary boundary is shown in the singular character of the formula (40). It turned out that the selection of the corresponding parameter value leads to a good approximation of the experimental data.

\acknowledgments{ 
The authors thank A.\,Y. Vul for attention to the work, M.\,G. Glazov for helpful comments, A.\,S. Khorolskaya for fruitful discussions. A.P. Meilakhs thanks RSF (grant 18-72-00131) for support.
}

\end{document}